\documentstyle[12pt,prl,aps,epsfig] {revtex}
\renewcommand{\baselinestretch}{1.6}
\newcommand{\be}{\begin{equation}}
\newcommand{\ee}{\end{equation}}

\newcommand{\bea}{\begin{eqnarray}}
\newcommand{\eea}{\end{eqnarray}}
\newcommand{\om}{\omega}
\newcommand{\Om}{\Omega}

\renewcommand{\baselinestretch}{1.2}
\begin{document}
\title{Improving the sensitivity of FM spectroscopy
using nano-mechanical cantilevers}
\author{B.M. Chernobrod, G.P. Berman, and P.W. Milonni \\
\small{Theoretical Division, Los Alamos National Laboratory, Los
Alamos, New Mexico 87545}} \maketitle
\renewcommand{\baselinestretch}{2}
\vspace{5mm}
\begin{abstract}
{It is suggested that nano-mechanical cantilevers can be employed
as  high-$Q$ filters to circumvent laser noise limitations on the
sensitivity of frequency modulation spectroscopy. In this approach
a cantilever is actuated by the radiation pressure of the
amplitude modulated light that emerges from an absorber. Numerical
estimates indicate that laser intensity noise will not prevent a
cantilever from operating in the thermal noise limit, where the
high $Q$'s of  cantilevers are most advantageous}.
\end{abstract}
\vspace{3mm} Frequency modulation spectroscopy (FMS)
\cite{bjorklund,song} has proven to be one of the most sensitive
absorption-based spectroscopic techniques.  The essential idea is
that when a frequency modulated laser beam enters an absorption
cell the emerging, partially absorbed  beam is amplitude modulated
(AM). If the modulation index of the frequency modulation is
sufficiently small, the spectrum of the incident FM beam consists
primarily of the peak at the carrier frequency $\om_c$ and
sidebands at $\om_c\pm\Om$, where $\Om$ is the modulation
frequency. In addition to the component at the carrier frequency,
the intensity of the output beam has a component that oscillates
sinusoidally at $\Om$ with an amplitude proportional to the
modulation index and to the difference in the attenuation
coefficients of the absorber at the frequencies $\om_c+\Om$ and
$\om_c-\Om$. The output of a photodetector is electronically
filtered and amplified, and the signal of interest that oscillates
at $\Om$ is extracted by a mixer. An absorption spectrum is
obtained by scanning the laser frequency over the spectral feature
of interest.

It is important for FMS that there be little spectral overlap
between the carrier and the sidebands. If the modulation frequency
is not high enough, the spectral wings of the sidebands and the
carrier will overlap, making the exact amplitude and phase balance
required for full FM beat cancellation impossible. Thus, one of
the major limiting factors in FMS is laser noise, which requires
that the modulation frequency be large compared with the laser
bandwidth. For laser bandwidths of 10-100 MHz, for instance,
modulation frequencies in the 100-1000 MHz range are desirable.
However, electronic detection systems, consisting of a photodetector and
several amplification cascades, produce an additional noise which
increases with increasing frequency, so that the shift to higher
modulation frequency could be inefficient. The highest sensitivity
of FMS is usually achieved with well stabilized, low-power
semiconductor lasers. While it is possible to substantially
reduce the laser technical noise and bandwidth in these
lasers, this is generally incompatible with the relatively high
laser powers required, for instance, for remote sensing or long-length
multipass cells.

In this letter we suggest the use of nano-mechanical cantilevers
(nano-resonators)  as filters with much higher $Q$ factors than
are currently possible by conventional methods. In this approach
the AM signal that emerges from the absorption cell (or is
backscattered in the case of remote sensing) actuates a cantilever
by resonant light pressure or by optical gradient forces. The
cantilever functions both as a high-frequency detector and as a
high-$Q$ filter. As discussed below, the use of cantilevers in FMS
in this way could offer the possibility of detecting molecules
with unprecedented sensitivity.

The lower limit on $\alpha L$, where $\alpha$ is the absorption coefficient and
$L$ is the total propagation length, can be estimated from the condition
that the signal-to-noise ratio ($SNR_c$) be unity. $SNR_c$ can be written as \be SNR_c={x_{\rm
sig}^2\over (x_{SN}^{\rm rms})^2 + (x_T^{\rm rms})^2 +(x_N^{\rm
rms})^2}, \ee
where $x_{\rm sig}$ is the vibrational amplitude of the cantilever,
$x_{SN}^{\rm rms}$ is the root mean square (rms) vibration noise
induced by the laser shot noise, $x_T^{\rm rms}$ is the rms
vibrational thermal noise, and $x_N^{\rm rms}$ is the vibrational noise
induced by the laser intensity noise. Substitution in (1) of the
expressions for the vibrational amplitudes gives \be SNR_c =
{Q^2(1+R)^2(\alpha L)^2P^2_0\over 4k_BTkc^2+Q(1+R)^2\om_0[P_0
\hbar\om + P_N(\om_0)^2]}, \ee
where $R$ is the reflection coefficient, $k$ is the spring
constant, $k_B$ is Boltzmann's constant, $T$ is the temperature,
$P_0$ is the incident laser power, $\om_0$ is the fundamental
frequency of the cantilever, and $P_N(\om)=\xi (\om) P_0$ is the
laser intensity noise, where $\xi (\om)$ is the relative intensity
noise (RIN). Consider as an example the following parameter
values: $T=4$ K, $k=0.3$ N/m, $Q=2\times 10^5$, $R=0.5$, $P_0=100$
$\mu$W, and $\om_0=20$ MHz. In this case the laser noise dominates
if the RIN satisfies the inequality $\xi (\om)> 1.8\times 10^{-5}
\ {\rm Hz}^{-1/2}$. This value of the RIN is typical for solid
state lasers \cite{tac}. Neglecting the shot noise and thermal
noise, we obtain $SNR_c \sim (\alpha L)^2/(\om_0\xi^2 (\om))$. The
condition $SNR_c = 1$ then gives $(\alpha L)_{\rm cantilever} =
\xi (\om_0)\sqrt {\om_0 / Q} =1.8\times 10^{-4}$.

Let us compare this estimate of the minimal $\alpha L$ with the
sensitivity of conventional electronic detection. The
photodetection usually involves at least three electronic stages
\cite{silver}. The first stage is the photodetector and
preamplifier or the photomultiplier, or avalanche photodiode; the
second stage is the lock-in-amplifier; and the third stage is the
output amplifier. Each stage produces noise. The noise of the two
first stages increases significantly at higher modulation
frequencies. We can characterize the noise by the noise figure $NF
=10\log [SNR_{\rm out}/SNR_{\rm in}]$ (where $SNR_{in}$ and
$SNR_{out}$ are the $SNR$ in the input and output signals,
correspondingly). For a modulation frequency greater than 10 MHz
the noise figure for the photodetector plus preamplifier is about
$NF_1 = 2$ dB; for the lock-in-amplifier $NF_2 = 3$ dB - 5 dB; and
for the output amplifier $NF_3 = 4$ dB. Thus the total noise
figure is $ NF = NF_1\times NF_2\times NF_3 \sim 10$ dB. The $SNR$
for electronic photodetection can be written as
 \be SNR_e =
{{g^2P^2_0(\alpha L)^2}\over {\Delta f e (g P_0 + 2k_B T/R) +
\Delta fg^2 \xi^2_{eff}(\om_0)P^2_0}}, \ee
 where $g = e\eta /\hbar\om$ . The
symbols in this equation are electric charge $e$, detector quantum
efficiency $\eta$, photon energy $\hbar\om$, bandwidth $\Delta f$,
resistance $R$, and $\xi_{eff} = \xi (\om)\times NF$. The first
term in the denominator corresponds to the laser shot noise, the
second term to thermal noise, and the third term to the laser
intensity noise. For the parameter values $R = 50 \ \Om$, $\eta =
0.8 $, $\xi(\om) = 1.8\times 10^{-5} \ {\rm Hz}^{-1/2}$, and $NF =
10$dB, the laser intensity noise dominates, and the condition
$SNR_c = 1$ for the minimum detectable absorption gives $(\alpha
L)_{\rm electronic} = \xi (\om) NF \Delta f^{1/2}$.

To avoid additional loss of signal, the
effective bandwidth $\Delta f$ should exceed the bandwidth of
modulation. For the modulation frequency $\om_m /2\pi = 20$ GHz and
the (highest) quality factor $Q_m = 2\times 10^8 $, the bandwidth
 $\Delta f= 100$ Hz. The assumption that the cantilever bandwidth $\om_0/Q=\Delta f $ implies $(\alpha L) _{\rm cantilever} =
0.1(\alpha L)_{\rm electronic}$, i.e., the smallest measurable absorption using the cantilever
 is less than the corresponding value for electronic
detection by at least an order of magnitude.

By choosing the cantilever resonance frequency appropriately, the
proposed sensor can be made to operate in the thermal noise limit.
To see this, let us assume a  laser noise spectrum $ P_N(\om)=
P_0\xi\left[{\Gamma^2/(\Gamma^2+(\om_L - \om )^2)}\right]^{1/2}$,
where $\xi $ is the spectral density of relative intensity noise
(RIN) at the center of the spectral distribution around the peak
of laser intensity noise at the frequency $\om_L$. If the
cantilever is to operate in the thermal noise limit the following
condition must be satisfied: $ x_T^{\rm rms}>x_N^{\rm rms}$, or
\be \Bigg[{{\mu}\over{1+\mu^2}}\Bigg]^{1/2}\xi <{c\over
(1+R)P_0}\sqrt{{4\pi k_BTk\over Q\Gamma}} \ee where $\mu\equiv
(\om_0 -\om_L)/\Gamma$. For $\om_L = 0.3$ MHz, $\Gamma = 1$ MHz,
and for the other parameters assumed above, the inequality (4)
gives $\mu > 5$, or $\om_0 > 5$ MHz.

We have assumed that radiation pressure rather than the
photothermal effect produces the dominant force in exciting the
cantilever vibrations. Experimental evidence  suggests that this
is indeed the case for Si cantilevers \cite{marti,yang}; namely, the
fact that the cantilever could be actuated at very high
temperatures,  where thermal gradients are much smaller than the
surface temperature, suggests that photothermal effects are
relatively small compared with radiation pressure.

Large resonance frequencies $\om_0$, and therefore small cantilevers,
are generally desirable for increasing the sensitivity.
If the cantilever is smaller than the spot size of
the beam incident upon it, actuation of the cantilever vibrations
may be inefficient. In this case one could use a scheme of
apertureless near-field microscopy. The tip is put in close
proximity to the cantilever surface, and focused light illiminates
the tip-surface region. The light intensity near the tip apex
exceeds the external intensity by an enhancement factor that could
be $\sim 10^6$ in the case of a plasmon resonance with a metallic
tip. Near the tip the field is very inhomogeneous, implying that
the gradient force could exceed the force of light pressure,
depending on the geometry.

\begin{figure}[t]
\centerline{\epsfig{figure=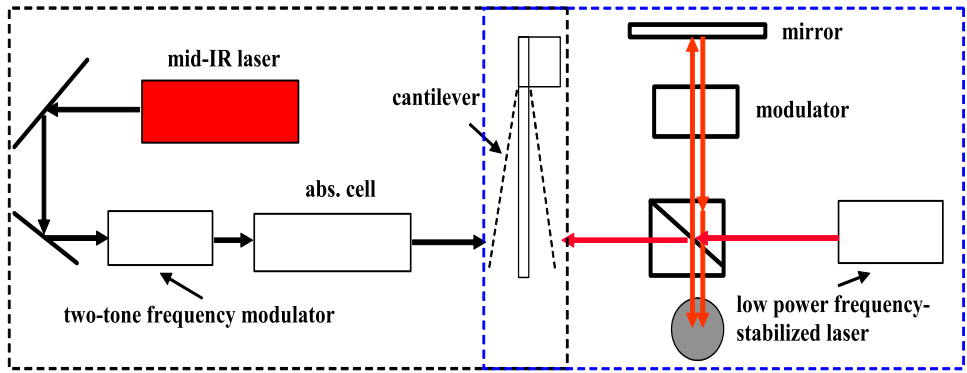,width=14.5cm,height=6.7cm,clip=}}
\caption{A frequency modulated laser beam  is passed through an
absorption cell and causes a cantilever to vibrate near its
resonant frequency. The cantilever vibrations are detected
interferometrically, as indicated in the right side of the
figure.} \label{fig1}
\end{figure}

Optical actuation of cantilevers by light pressure has been
demonstrated \cite{marti,yang}. In the experiments of Yang {\it et
al.}, deflections of a $60\times 6$ $\mu$m$^2$ cantilever with
680-nm, 40-$\mu$W laser radiation of beam size $\sim 300\times
100$ $\mu$m$^2$ were observed. Deflections were observed for
temperatures as high as 780 C. Their cantilever had a $Q$ factor
$\approx 10^5$ and a spring constant $k=4.4\times 10^{-3}$ N/m.
For these parameters we estimate a force sensitivity $F_T\approx
10^{-16}$ N, in rough agreement with the value quoted by Yang {\it
et al.} The condition (4) for the cantilever to operate in the
thermal noise limit is found to be easily realized for these
parameters. Such numerical estimates support the viability of the
proposed cantilever sensor. Note also that in Reference
\cite{zala} a cantilever quality factor up to $Q\approx 10^5$ was
achieved for driven oscillations of the cantilever for a laser
power of a few hundred microwatts.

In conclusion, we have presented estimates indicating that
nano-mechanical cantilevers can be employed as high-$Q$ filters to
circumvent laser noise limitations on the sensitivity of frequency
modulation spectroscopy.

\section*{Acknowledgments}

We thank Professor Umar Mohideen and Dr. C. E. Strauss for helpful
discussions. This work was supported by the Department of Energy
under the contract W-7405-ENG-36 and DOE Office of Basic Energy
Sciences, and by the Defense Advanced Research Projects Agency.

\end{document}